\documentclass[11pt,a4paper]{article}
\usepackage[T1]{fontenc}
\usepackage[utf8]{inputenc}
\usepackage{lmodern}
\usepackage[margin=25mm]{geometry}
\usepackage{amsmath,amssymb,amsthm,bm,mathtools,mathrsfs}
\usepackage{authblk,booktabs,array,longtable}
\usepackage{microtype}
\usepackage[numbers,sort&compress]{natbib}
\usepackage[colorlinks=true,linkcolor=blue,citecolor=blue,urlcolor=blue]{hyperref}
\hypersetup{pdftitle={Fermions and Superradiance: Expanded Derivations},
pdfauthor={Wen-Xiang Chen and Yao-Guang Zheng},
pdfsubject={Expanded and revised theoretical-physics manuscript}}
\usepackage{enumitem}
\setlist{nosep}
\allowdisplaybreaks[2]
\numberwithin{equation}{section}
\newcommand{\dd}{\mathrm{d}}
\newcommand{\ii}{\mathrm{i}}
\newcommand{\ee}{\mathrm{e}}

\newcommand{\diag}{\operatorname{diag}}

\newcommand{\Refl}{\mathscr{R}}
\newcommand{\Trans}{\mathscr{T}}
\newcommand{\cs}{c_{\mathrm{s}}}
\newcommand{\rH}{r_{\mathrm{H}}}
\newcommand{\OH}{\Omega_{\mathrm{H}}}

\title{Fermions can also produce super-radiation phenomena }
\author[1]{Wen-Xiang Chen\\wxchen4277@qq.com}
\affil[1]{School of Electronic Information, Guangzhou City University of Technology,\\
Guangzhou 510800, China}

\begin{document}
\maketitle
\begin{abstract}
Can a system containing fermions display superradiant phenomena without contradicting
the absence of classical superradiant scattering for the minimally coupled Dirac
field? We address this question by distinguishing three scattering and emission
problems. First, we derive the Dirac continuity equation and its flux-normalized
reflection bound, solve the massless electrostatic problem directly, and work out
massive potential-step matching on both positive- and negative-energy branches.
These calculations show why a prescribed real potential or boundary selection
cannot, by itself, reverse the Dirac flux identity. Second, we derive the
fermionic Bogoliubov normalization and the corresponding particle-production
formula, which permit spontaneous quantum emission while preserving Pauli
blocking. Third, under a dilute molecular-pair and long-wavelength approximation,
we construct a neutral pair-condensate effective theory, derive its collective
wave equation, and obtain the amplification condition for a rotating draining
flow. A reproducible numerical example illustrates this conditional collective
mechanism. We also supply the intermediate geometric steps of the Hasimoto
transformation and explain why its focusing nonlinear Schr\"odinger equation
cannot simply be identified with the Dirac equation or with a repulsive
pair-condensate equation. The result is a consistent separation of classical
Dirac scattering, quantum fermion emission, and bosonic collective amplification
in a system built from fermionic constituents.
\end{abstract}
\noindent\textbf{Keywords:} superradiance; Dirac current; fermion pairs;
Bogoliubov transformation; acoustic geometry; Hasimoto transformation

\newpage
\tableofcontents
\newpage

\section{Introduction}
\label{sec:introduction}

The word superradiance describes more than one physical mechanism. In the
collective optical effect introduced by Dicke, the relevant radiation is emitted
coherently by an ensemble of emitters \cite{Dicke1954}. In rotational or charged
wave scattering, the defining observation is instead an outgoing energy flux
larger than the incident energy flux, with the difference supplied by the
background. A reflecting environment can feed this gain back into an instability,
as in the black-hole-bomb proposal \cite{PressTeukolsky1972}. Energy and charge
extraction by black holes provide the corresponding relativistic setting
\cite{Bekenstein1973}.

The distinction between the field being scattered and the constituents of its
medium is essential. A classical, minimally coupled Dirac wave does not acquire
the usual superradiant reflection gain in the standard Kerr scattering problem
\cite{Unruh1973}. Nevertheless, quantized fermionic fields can exhibit spontaneous
emission on black-hole backgrounds \cite{Unruh1974,Hawking1975}. More recently,
quantum fermion superradiance on charged black holes has been studied with an
explicit treatment of vacuum choices \cite{AlvarezWinstanley2025}. These quantum
statements do not assert that the classical Dirac reflection coefficient exceeds
unity.

There is a second distinction. Two fermions can form a composite boson, and a
condensate of such pairs has collective excitations whose effective equations
differ from the one-particle Dirac equation. Pairing is established in the Cooper
and Bardeen--Cooper--Schrieffer frameworks \cite{Cooper1956,BCS1957}. For strongly
bound, dilute pairs, the molecular Bose-condensate regime has a microscopic and
experimental basis \cite{PieriStrinati2003,Petrov2004,Jochim2003,Greiner2003}.
If an appropriate moving pair condensate is realized, its sound modes can then
be considered within analogue-gravity scattering theory. This is a conditional
route to collective amplification, rather than an exception to the Dirac-current
argument.

The supplied earlier discussion \cite{ChenZheng2023} raises the useful question
of how prescribed boundary conditions, fermion pairing, and vortex geometry
affect superradiance. Its argument, however, requires a separation of the
conserved quantities involved. A scalar Wronskian cannot replace the Dirac
probability current; an uncertainty relation for measurement disturbance cannot
be used as a source term in a continuity equation; and the geometric Hasimoto
field cannot be equated with a fermionic wave function without an additional
dynamical construction. The following sections make these distinctions explicit
and provide the missing intermediate calculations.

Our purpose is therefore a derivational reconstruction with clearly stated
assumptions. The standard no-amplification result, the existence of quantum
fermion emission, the emergence of molecular bosons, and acoustic superradiance
are established ingredients. The conditional pair-condensate example below
combines these ingredients to specify a calculable interpretation of
``superradiant phenomena in a fermionic system.'' It does not establish a new
microscopic black-hole solution, a new violation of the Dirac reflection bound,
or an experimentally demonstrated amplification mechanism in a Fermi gas.
For broader background on the scattering terminology, see
Ref.~\cite{BritoCardosoPani2020}.

\section{Definitions, conventions, and scope}
\label{sec:definitions}

\subsection{Three observables that must be kept separate}

Let $F_{\mathrm{in}}>0$ be an incoming flux and let
$F_{\mathrm{ref}}>0$ denote the magnitude of the reflected outgoing flux.
We define
\begin{equation}
 \Refl=\frac{F_{\mathrm{ref}}}{F_{\mathrm{in}}},\qquad
 Z=\Refl-1.
 \label{eq:gain-definition}
\end{equation}
Classical scattering amplification means $Z>0$. A ratio of wave amplitudes is
equal to this flux ratio only when the asymptotic mode normalizations and
velocities agree. Spontaneous quantum emission is instead characterized by
\begin{equation}
 N_{\mathrm{out}}=
 \langle 0,\mathrm{in}|\hat a_{\mathrm{out}}^\dagger
 \hat a_{\mathrm{out}}|0,\mathrm{in}\rangle>0.
 \label{eq:spontaneous-definition}
\end{equation}
There need not be an incident classical wave in
Eq.~\eqref{eq:spontaneous-definition}. A dynamical instability is a third
statement: a homogeneous initial-boundary-value problem admits a mode
$\ee^{-\ii\omega t}$ with $\operatorname{Im}\omega>0$.
Neither a nonzero $N_{\mathrm{out}}$ nor a single-pass gain alone proves such
an instability.

\subsection{Conventions and assumptions}

We use signature $(+,-,-,-)$ for the relativistic discussion and natural units
$\hbar=c=1$ in Secs.~\ref{sec:dirac}--\ref{sec:quantum}. In the nonrelativistic
pair-condensate sections, $\hbar$ is restored. The sound speed $\cs$ is unrelated
to the vacuum speed of light. The fermion mass is $m_{\mathrm f}$ and the pair
mass is $M_{\mathrm B}=2m_{\mathrm f}$. An integer $m$ appearing in a rotating
mode is an azimuthal number, not a mass. The real electrostatic potential energy
is $V=eA_0$ in the flat-space model. The Dirac number current and electric
current are different quantities:
\begin{equation}
 j^\mu=\bar\Psi\gamma^\mu\Psi,\qquad J^\mu=e j^\mu.
 \label{eq:number-charge}
\end{equation}
It is $j^0$, rather than $J^0$ for a charge of either sign, that has the
positive-density interpretation.

For the classical Dirac bound we assume a linear Hermitian bulk equation, a
real prescribed background, one incident channel, and no independent source at
the outgoing boundary. For quantum emission we specify distinct in and out
particle bases; their choice is part of the physical state specification. For
the pair example we assume neutral, dilute bound molecules with positive
effective compressibility, a smooth stationary flow outside an absorbing core,
and wavelengths long compared with the healing length. These are distinct
models; a conclusion from one is not automatically a theorem about another.

\section{Classical Dirac scattering: the flux bound and an exact example}
\label{sec:dirac}

\subsection{Deriving the continuity equation}

\paragraph{Step 1: choose a consistent two-component Hamiltonian.}
For the one-dimensional problem take
\begin{equation}
 \ii\partial_t\Psi=H\Psi,\qquad
 H=-\ii\sigma_x\partial_x+m_{\mathrm f}\sigma_z+V(x)\mathbf 1,
 \label{eq:dirac-hamiltonian}
\end{equation}
where $\sigma_x^2=\sigma_z^2=\mathbf 1$ and
$\{\sigma_x,\sigma_z\}=0$. This is the Hamiltonian representation associated
with $\gamma^0=\sigma_z$ and $\gamma^1=\ii\sigma_y$.
For real $V$, define
\begin{equation}
 \rho=\Psi^\dagger\Psi,\qquad j=\Psi^\dagger\sigma_x\Psi.
 \label{eq:dirac-rho-j}
\end{equation}

\paragraph{Step 2: take the time derivative of the density.}
Equation~\eqref{eq:dirac-hamiltonian} and its adjoint give
\begin{align}
 \partial_t\Psi&=-\sigma_x\partial_x\Psi
 -\ii(m_{\mathrm f}\sigma_z+V)\Psi,\\
 \partial_t\Psi^\dagger&=-(\partial_x\Psi^\dagger)\sigma_x
 +\ii\Psi^\dagger(m_{\mathrm f}\sigma_z+V).
\end{align}
Substitute these expressions into the product rule:
\begin{align}
 \partial_t\rho
 &= (\partial_t\Psi^\dagger)\Psi+\Psi^\dagger\partial_t\Psi\nonumber\\
 &=-(\partial_x\Psi^\dagger)\sigma_x\Psi
   -\Psi^\dagger\sigma_x\partial_x\Psi\nonumber\\
 &=-\partial_x(\Psi^\dagger\sigma_x\Psi).
\end{align}
The mass and real-potential contributions cancel separately. Thus
\begin{equation}
 \boxed{\partial_t\rho+\partial_x j=0.}
 \label{eq:continuity}
\end{equation}
For a stationary mode $\Psi=\ee^{-\ii\omega t}\chi(x)$, $\rho$ is independent
of time and $\dd j/\dd x=0$.

\paragraph{Step 3: evaluate the conserved current in asymptotic channels.}
Let a mode of amplitude $A$ and group velocity $v$ carry current $v|A|^2$
in a density-normalized convention. If $v_{\mathrm L}>0$ is the magnitude of
the left asymptotic velocity and $v_{\mathrm R}>0$ the transmitted velocity,
then a one-sided experiment has
\begin{equation}
 j_{\mathrm L}=v_{\mathrm L}(|\mathcal I|^2-|\mathcal R|^2),\qquad
 j_{\mathrm R}=v_{\mathrm R}|\mathcal T|^2.
\end{equation}
Equating these currents gives
\begin{equation}
 \Refl_{\mathrm D}=\frac{|\mathcal R|^2}{|\mathcal I|^2},\qquad
 \Trans_{\mathrm D}=\frac{v_{\mathrm R}}{v_{\mathrm L}}
 \frac{|\mathcal T|^2}{|\mathcal I|^2},\qquad
 \boxed{\Refl_{\mathrm D}+\Trans_{\mathrm D}=1.}
 \label{eq:dirac-unitarity}
\end{equation}
Since the transmitted channel is outgoing to the right,
$\Trans_{\mathrm D}\geq0$ and $0\leq\Refl_{\mathrm D}\leq1$.
Selecting the outgoing channel by the sign of its momentum alone can be
incorrect when its kinetic energy is negative; the velocity or current decides
its propagation direction.

\subsection{The massless equations appearing in the original model}
\label{subsec:massless}

\paragraph{Step 1: write the component system.}
Set $m_{\mathrm f}=0$ and $\chi=(f_1,f_2)^{\mathsf T}$. The stationary
equation is
\begin{equation}
 \frac{\dd f_1}{\dd x}=\ii\epsilon(x)f_2,\qquad
 \frac{\dd f_2}{\dd x}=\ii\epsilon(x)f_1,\qquad
 \epsilon(x)=\omega-V(x).
 \label{eq:original-system}
\end{equation}
Its current is
\begin{equation}
 j=f_1^*f_2+f_2^*f_1=2\operatorname{Re}(f_1^*f_2).
 \label{eq:component-current}
\end{equation}
Direct differentiation using Eq.~\eqref{eq:original-system} gives
\begin{align}
 \frac{\dd}{\dd x}(f_1^*f_2)
 &=-\ii\epsilon|f_2|^2+\ii\epsilon|f_1|^2,\\
 \frac{\dd}{\dd x}(f_2^*f_1)
 &=-\ii\epsilon|f_1|^2+\ii\epsilon|f_2|^2.
\end{align}
Their sum is zero for every real $\epsilon$.

\paragraph{Step 2: diagonalize the two first-order equations.}
Introduce
\begin{equation}
 h_+=f_1+f_2,\qquad h_-=f_1-f_2.
\end{equation}
Adding and subtracting Eq.~\eqref{eq:original-system}, respectively, yields
\begin{equation}
 h_+'=\ii\epsilon h_+,\qquad h_-'=-\ii\epsilon h_-.
\end{equation}
For a locally integrable real potential, their exact solutions are
\begin{equation}
 h_\pm(x)=C_\pm\exp\left[\pm\ii
 \int_{x_0}^{x}(\omega-V(u))\,\dd u\right].
 \label{eq:chiral-solutions}
\end{equation}
The two chiralities do not couple. In particular, there is no singularity in
this first-order solution when $\epsilon$ changes sign.

\paragraph{Step 3: impose one-sided incidence.}
On the left, write the asymptotic amplitudes as
\begin{equation}
 (f_1,f_2)=\mathcal I(1,1)\ee^{\ii k_{\mathrm L}x}
 +\mathcal R(1,-1)\ee^{-\ii k_{\mathrm L}x}.
\end{equation}
The respective currents are $2|\mathcal I|^2$ and
$-2|\mathcal R|^2$. A right-moving transmitted wave on the right has
$f_1=f_2$, hence $h_-=0$. Equation~\eqref{eq:chiral-solutions} then implies
$C_-=0$ throughout the domain, including the left asymptotic region.
Therefore
\begin{equation}
 \boxed{\mathcal R=0,\qquad\Refl_{\mathrm D}=0,
 \qquad\Trans_{\mathrm D}=1}
 \label{eq:massless-transparent}
\end{equation}
for this strictly one-dimensional massless electrostatic model, with no
chirality-mixing interaction or boundary source. This stronger result is
specific to the model; angular barriers, masses, or additional couplings can
produce reflection in other Dirac scattering problems.

\paragraph{Step 4: examine a frequency-parametrized potential.}
For $V=-y\omega$, one simply has
\begin{equation}
 \epsilon=(1+y)\omega.
\end{equation}
This changes the phase in Eq.~\eqref{eq:chiral-solutions}; it does not change
the continuity equation. Even a large real $y$ cannot turn
Eq.~\eqref{eq:dirac-unitarity} into an amplification relation. If the potential
is chosen separately for each frequency, the family describes different
backgrounds unless a dispersive medium supplies that dependence. Such a medium
requires its own degrees of freedom and flux accounting.

\subsection{Massive step: matching both kinetic-energy branches}
\label{subsec:massive-step}

\paragraph{Step 1: solve the constant-potential eigenvalue equation.}
For $\chi=u(p)\ee^{\ii px}$ and $\epsilon=\omega-V$, the stationary equation
becomes
\begin{equation}
 \begin{pmatrix}m_{\mathrm f}&p\\p&-m_{\mathrm f}\end{pmatrix}
 \begin{pmatrix}u_1\\u_2\end{pmatrix}
 =\epsilon\begin{pmatrix}u_1\\u_2\end{pmatrix}.
\end{equation}
The characteristic polynomial gives
\begin{equation}
 \epsilon^2=p^2+m_{\mathrm f}^2.
\end{equation}
Taking $u_1=1$ away from the threshold $\epsilon=-m_{\mathrm f}$, the second
row gives
\begin{equation}
 u(p)=\begin{pmatrix}1\\q\end{pmatrix},\qquad
 q=\frac{p}{\epsilon+m_{\mathrm f}},\qquad
 j[u]=2q,\qquad\rho[u]=1+q^2.
 \label{eq:step-spinor}
\end{equation}
The current-to-density ratio is
\begin{equation}
 \frac{j}{\rho}=\frac{2p(\epsilon+m_{\mathrm f})}
 {(\epsilon+m_{\mathrm f})^2+p^2}
 =\frac{p}{\epsilon}=\frac{\dd\epsilon}{\dd p}.
 \label{eq:step-velocity}
\end{equation}
For the middle equality we used $p^2=\epsilon^2-m_{\mathrm f}^2$.

\paragraph{Step 2: choose the outgoing transmitted root.}
Take $V=0$ for $x<0$ and $V=V_0$ for $x>0$, with $\omega>m_{\mathrm f}$.
Let
\begin{equation}
 p_{\mathrm L}=\sqrt{\omega^2-m_{\mathrm f}^2},\qquad
 q_{\mathrm L}=\frac{p_{\mathrm L}}{\omega+m_{\mathrm f}}>0.
\end{equation}
If $\epsilon_{\mathrm R}=\omega-V_0$ is propagating,
$|\epsilon_{\mathrm R}|>m_{\mathrm f}$, choose
\begin{equation}
 p_{\mathrm R}=\operatorname{sgn}(\epsilon_{\mathrm R})
 \sqrt{\epsilon_{\mathrm R}^2-m_{\mathrm f}^2},\qquad
 q_{\mathrm R}=\frac{p_{\mathrm R}}{\epsilon_{\mathrm R}+m_{\mathrm f}}>0.
 \label{eq:outgoing-root}
\end{equation}
Thus $p_{\mathrm R}<0$ on the negative-energy branch, but
$p_{\mathrm R}/\epsilon_{\mathrm R}>0$: the mode still travels to the right.
Thresholds are obtained by limits and are excluded from the displayed spinor
parametrization.

\paragraph{Step 3: impose spinor continuity at the finite step.}
The wave functions are
\begin{align}
 \chi_{\mathrm L}&=\mathcal I
 \begin{pmatrix}1\\q_{\mathrm L}\end{pmatrix}\ee^{\ii p_{\mathrm L}x}
 +\mathcal R\begin{pmatrix}1\\-q_{\mathrm L}\end{pmatrix}
 \ee^{-\ii p_{\mathrm L}x},\\
 \chi_{\mathrm R}&=\mathcal T
 \begin{pmatrix}1\\q_{\mathrm R}\end{pmatrix}\ee^{\ii p_{\mathrm R}x}.
\end{align}
Integrating the first-order equation over an infinitesimal interval around a
finite step leaves no delta-function term, so both components are continuous:
\begin{equation}
 \mathcal I+\mathcal R=\mathcal T,\qquad
 q_{\mathrm L}(\mathcal I-\mathcal R)=q_{\mathrm R}\mathcal T.
\end{equation}
Eliminating $\mathcal T$ and then $\mathcal R$ gives
\begin{equation}
 \frac{\mathcal R}{\mathcal I}=
 \frac{q_{\mathrm L}-q_{\mathrm R}}{q_{\mathrm L}+q_{\mathrm R}},\qquad
 \frac{\mathcal T}{\mathcal I}=
 \frac{2q_{\mathrm L}}{q_{\mathrm L}+q_{\mathrm R}}.
\end{equation}

\paragraph{Step 4: normalize by current, rather than by component amplitude.}
Using $j=2q$ per unit component amplitude, we obtain
\begin{align}
 \Refl_{\mathrm D}&=
 \frac{(q_{\mathrm L}-q_{\mathrm R})^2}
 {(q_{\mathrm L}+q_{\mathrm R})^2},\\
 \Trans_{\mathrm D}&=
 \frac{q_{\mathrm R}}{q_{\mathrm L}}
 \left|\frac{\mathcal T}{\mathcal I}\right|^2
 =\frac{4q_{\mathrm L}q_{\mathrm R}}
 {(q_{\mathrm L}+q_{\mathrm R})^2}.
 \label{eq:massive-step-flux}
\end{align}
The numerator of their sum is
$(q_{\mathrm L}-q_{\mathrm R})^2+4q_{\mathrm L}q_{\mathrm R}
=(q_{\mathrm L}+q_{\mathrm R})^2$, verifying
Eq.~\eqref{eq:dirac-unitarity} on either propagating branch.
If $|\epsilon_{\mathrm R}|<m_{\mathrm f}$, the decaying solution carries zero
asymptotic current for a semi-infinite step and $\Refl_{\mathrm D}=1$.
These are stationary one-particle identities. Interpreting the negative-energy
sector as particles and antiparticles requires the additional vacuum
specification of Sec.~\ref{sec:quantum}.

\section{Why the scalar Wronskian gives a different identity}
\label{sec:scalar}

\subsection{Derivation for a massless charged scalar}

For comparison, consider
\begin{equation}
 [(\partial_t+\ii V)^2-\partial_x^2]\Phi=0,\qquad
 \Phi=\ee^{-\ii\omega t}f(x).
\end{equation}
The radial equation is
\begin{equation}
 f''+(\omega-V)^2f=0.
 \label{eq:scalar-wave}
\end{equation}
Because its coefficient is real, both $f$ and $f^*$ solve the same equation.
For this second-order scalar equation define
\begin{equation}
 W[f^*,f]=f^*f'-f(f^*)',\qquad
 j_{\mathrm{KG}}=\frac{W}{2\ii}.
\end{equation}
Differentiating and using Eq.~\eqref{eq:scalar-wave},
\begin{equation}
 W'=f^*f''-f(f^*)''=
 - (\omega-V)^2f^*f+(\omega-V)^2ff^*=0.
\end{equation}
For a constant-potential mode $f=A\ee^{\ii kx}$, this convention gives
\begin{equation}
 j_{\mathrm{KG}}=k|A|^2,\qquad
 \rho_{\mathrm{KG}}=(\omega-V)|A|^2.
\end{equation}
Unlike the Dirac density, the scalar density is not positive on all branches.

For a step with $V_{\mathrm L}=0$, take $k_{\mathrm L}=\omega>0$ and a
right-moving massless transmitted branch with
$k_{\mathrm R}=\omega-V_0$. Its group velocity is
$k_{\mathrm R}/(\omega-V_0)=1$, including when both numerator and denominator
are negative. The asymptotic Wronskians give
\begin{equation}
 \omega(|\mathcal I|^2-|\mathcal R|^2)
 = (\omega-V_0)|\mathcal T|^2,
\end{equation}
or
\begin{equation}
 \boxed{\frac{|\mathcal R|^2}{|\mathcal I|^2}
 =1-\frac{\omega-V_0}{\omega}
 \frac{|\mathcal T|^2}{|\mathcal I|^2}.}
 \label{eq:scalar-amplification}
\end{equation}
The transmitted scalar norm flux is negative for $0<\omega<V_0$.
Consequently, nonzero transmission gives reflected amplification.
At the incident asymptotic boundary, where $V=0$ and the incident and reflected
frequencies agree, the reflected-to-incident energy-flux ratio is also
$|\mathcal R/\mathcal I|^2$. Global charged-field energy accounting includes
the background that sustains the potential.

\subsection{Why this calculation cannot be inserted into the Dirac problem}

Eliminating $f_2$ from Eq.~\eqref{eq:original-system}, where
$\epsilon\ne0$, gives
\begin{align}
 f_1''&=\ii\epsilon'f_2+\ii\epsilon f_2'
 =\frac{\epsilon'}{\epsilon}f_1'-\epsilon^2f_1,\\
 f_1''-\frac{\epsilon'}{\epsilon}f_1'+\epsilon^2f_1&=0.
 \label{eq:dirac-eliminated}
\end{align}
This is not Eq.~\eqref{eq:scalar-wave}. In particular, for two solutions of
Eq.~\eqref{eq:dirac-eliminated}, Abel's identity gives
\begin{equation}
 W'=\frac{\epsilon'}{\epsilon}W,\qquad
 \left(\frac{W}{\epsilon}\right)'=0,
 \label{eq:abel}
\end{equation}
on an interval where division by $\epsilon$ is valid. With
$f_2=f_1'/(\ii\epsilon)$, the original current becomes
\begin{equation}
 j=\frac{f_1^*f_1'-f_1(f_1^*)'}{\ii\epsilon}
 =\frac{W[f_1^*,f_1]}{\ii\epsilon}.
 \label{eq:dirac-current-wronskian}
\end{equation}
It is the extra $1/\epsilon$ factor that restores the Dirac flux identity.
At a zero of $\epsilon$ one should return to the regular first-order system,
rather than use the divided equation. Also, complex conjugation of the complete
first-order system changes $\ii$ to $-\ii$; it cannot be assumed to produce
another solution with unchanged component equations and boundary data.

\section{Curved-spacetime current and the role of boundaries}
\label{sec:curved}

\subsection{Covariant current conservation}

Choose a tetrad with $\gamma^\mu=e^\mu{}_a\gamma^a$ and
$\{\gamma^\mu,\gamma^\nu\}=2g^{\mu\nu}$. For a real electromagnetic field,
\begin{equation}
 (\ii\gamma^\mu D_\mu-m_{\mathrm f})\Psi=0,\qquad
 D_\mu\Psi=(\nabla_\mu+\ii eA_\mu)\Psi.
\end{equation}
With $\bar\Psi=\Psi^\dagger\gamma^{\hat0}$, its adjoint is
\begin{equation}
 \ii(D_\mu\bar\Psi)\gamma^\mu+m_{\mathrm f}\bar\Psi=0,
\end{equation}
where the adjoint covariant derivative has the opposite electromagnetic sign.
Multiplying the first equation on the left by $\bar\Psi$ and the adjoint
equation on the right by $\Psi$, then adding, cancels the mass terms:
\begin{equation}
 \ii\bar\Psi\gamma^\mu D_\mu\Psi
 +\ii(D_\mu\bar\Psi)\gamma^\mu\Psi=0.
\end{equation}
Metric compatibility of the spin connection gives $\nabla_\mu\gamma^\nu=0$;
the electromagnetic terms cancel as well. Therefore
\begin{equation}
 \boxed{\nabla_\mu j^\mu=0,\qquad j^\mu=\bar\Psi\gamma^\mu\Psi.}
 \label{eq:covariant-current}
\end{equation}

\subsection{The horizon flux has the required sign}

In a local orthonormal frame,
\begin{equation}
 j^{\hat0}=\Psi^\dagger\Psi,\qquad
 j^{\hat i}=\Psi^\dagger\alpha^{\hat i}\Psi.
\end{equation}
For any unit spatial vector $\bm n$, the Hermitian matrix
$\bm n\cdot\bm\alpha$ has square $\mathbf1$ and eigenvalues $\pm1$.
Hence
\begin{equation}
 |\bm n\cdot\bm j|\leq j^{\hat0}.
\end{equation}
Maximizing over $\bm n$ gives $|\bm j|\leq j^{\hat0}$. Thus the classical
Dirac number current is future causal. If $\chi^\mu$ is a future-directed
null generator of the future horizon, then, in signature $(+,-,-,-)$,
\begin{equation}
 j_\mu\chi^\mu\geq0.
 \label{eq:horizon-positive}
\end{equation}
The integral of this contraction with the positive horizon measure is the
number flux entering the future horizon.

Integrate Eq.~\eqref{eq:covariant-current} over a stationary exterior region
between two time slices. The time-slice contributions cancel for a steady
scattering state. With $N_{\mathrm{in}}$, $N_{\mathrm{ref}}$, and
$N_{\mathrm H}$ defined as positive magnitudes in their physical directions,
the remaining boundary terms yield
\begin{equation}
 N_{\mathrm{in}}=N_{\mathrm{ref}}+N_{\mathrm H},\qquad
 N_{\mathrm H}\geq0.
 \label{eq:curved-balance}
\end{equation}
For elastic positive-frequency reflection at the same asymptotic boundary,
each incoming or reflected quantum carries the same energy $\omega$, so the
energy reflection ratio has the same bound. This argument assumes no extra
incoming channel from a past horizon or driven inner boundary.

The frequency measured along a charged rotating horizon is shifted by the
background rotation and electrostatic potential. In the convention used for
the asymptotic particle energy, write
\begin{equation}
 \widetilde\omega=\omega-m\OH-e\Phi_{\mathrm H}.
 \label{eq:shifted-frequency}
\end{equation}
The sign of $\widetilde\omega$ is relevant to particle-basis choices and to
bosonic energy extraction. It does not invalidate the positivity statement
\eqref{eq:horizon-positive} for the classical Dirac number current.

\subsection{Passive boundaries, driven boundaries, and postselection}

A self-adjoint reflecting wall has zero net normal number current. A passive
absorbing boundary removes nonnegative number flux. These boundary conditions
are compatible with Eq.~\eqref{eq:curved-balance}. They do not supply the
negative absorbed norm flux needed for ordinary scalar amplification.

For comparison, explicitly change the flat Hamiltonian to
$H=H_{\mathrm{Herm}}+\ii\Gamma(x)$, with real $\Gamma$. Repeating the density
derivative now gives, with $\hbar$ restored,
\begin{equation}
 \partial_t\rho+\nabla\cdot\bm j=\frac{2\Gamma}{\hbar}\rho.
 \label{eq:source-current}
\end{equation}
In a stationary one-dimensional domain this integrates to
\begin{equation}
 F_{\mathrm{ref}}+F_{\mathrm{trans}}-F_{\mathrm{in}}
 =\frac{2}{\hbar}\int\Gamma\rho\,\dd x.
\end{equation}
Gain is possible for an active medium, but its source is displayed on the
right-hand side. The equation differs from the source-free model.

Similarly, an independent incoming wave from the other boundary must be
included in the total incoming flux. A ratio formed using only one input can
exceed unity through interference without showing net gain. If a subset of
outcomes is postselected with success probability $p$, conditional expectations
are divided by $p$. That normalization does not create an unconditioned current
source. These observations explain how boundary manipulations can change a
measurement while leaving the local continuity identity intact.

\section{Quantum fermion emission and Pauli blocking}
\label{sec:quantum}

\subsection{Mode mixing and the normalization identity}

The classical-current proof does not fix the quantum vacuum. Expand the field
in complete in and out mode bases:
\begin{equation}
 \hat\Psi=\sum_\lambda
 (\hat a^{\mathrm{in}}_\lambda u^{\mathrm{in}}_\lambda
 +\hat b^{\mathrm{in}\dagger}_\lambda v^{\mathrm{in}}_\lambda)
 =\sum_\lambda
 (\hat a^{\mathrm{out}}_\lambda u^{\mathrm{out}}_\lambda
 +\hat b^{\mathrm{out}\dagger}_\lambda v^{\mathrm{out}}_\lambda).
\end{equation}
Here $\lambda$ labels normalized discrete modes or suitable wave packets; sums
are replaced by integrals for a continuum. A diagonal two-mode block has
\begin{equation}
 \hat a_{\mathrm{out}}=\alpha\hat a_{\mathrm{in}}
 +\beta\hat b_{\mathrm{in}}^\dagger,
 \label{eq:fermion-bogoliubov}
\end{equation}
with a complementary antiparticle transformation chosen to preserve all
canonical anticommutators. Using
$\{\hat a,\hat a^\dagger\}=\{\hat b,\hat b^\dagger\}=1$ and vanishing mixed
anticommutators, we find
\begin{align}
 1&=\{\hat a_{\mathrm{out}},\hat a_{\mathrm{out}}^\dagger\}\nonumber\\
 &=|\alpha|^2\{\hat a_{\mathrm{in}},\hat a_{\mathrm{in}}^\dagger\}
 +|\beta|^2\{\hat b_{\mathrm{in}}^\dagger,\hat b_{\mathrm{in}}\}\nonumber\\
 &=|\alpha|^2+|\beta|^2.
 \label{eq:fermionic-normalization}
\end{align}
The plus sign is a consequence of fermionic anticommutation.

\subsection{Particle production with an incident occupation}

For an input density operator with no particle--antiparticle coherence, define
$n_a=\langle\hat a_{\mathrm{in}}^\dagger\hat a_{\mathrm{in}}\rangle$ and
$n_b=\langle\hat b_{\mathrm{in}}^\dagger\hat b_{\mathrm{in}}\rangle$.
Multiplication of Eq.~\eqref{eq:fermion-bogoliubov} by its adjoint gives
\begin{align}
 \langle\hat a_{\mathrm{out}}^\dagger\hat a_{\mathrm{out}}\rangle
 &=|\alpha|^2 n_a+|\beta|^2
 \langle\hat b_{\mathrm{in}}\hat b_{\mathrm{in}}^\dagger\rangle\nonumber\\
 &=|\alpha|^2 n_a+|\beta|^2(1-n_b)\nonumber\\
 &=n_a+|\beta|^2(1-n_a-n_b).
 \label{eq:fermion-occupation}
\end{align}
We used $\hat b\hat b^\dagger=1-\hat b^\dagger\hat b$ and then
Eq.~\eqref{eq:fermionic-normalization}. For the in vacuum,
\begin{equation}
 N_{\mathrm{out}}=|\beta|^2,\qquad0\leq N_{\mathrm{out}}\leq1.
 \label{eq:fermion-spontaneous}
\end{equation}
This gives spontaneous emission with no incoming particle, while the factors
$1-n$ encode Pauli blocking. It is not the assertion
$\Refl_{\mathrm D}>1$. Even when the output occupation exceeds a small
input occupation, the difference can be the spontaneous contribution in
Eq.~\eqref{eq:fermion-occupation}, rather than a classical reflected gain.

For bosonic modes, the analogous commutator calculation gives instead
\begin{equation}
 |\alpha_{\mathrm B}|^2-|\beta_{\mathrm B}|^2=1,\qquad
 n_{a,\mathrm{out}}=|\alpha_{\mathrm B}|^2 n_a
 +|\beta_{\mathrm B}|^2(1+n_b).
 \label{eq:boson-occupation}
\end{equation}
The different algebra determines the different occupation factors. The
canonical coefficient $\alpha_{\mathrm B}$ in this formula need not be
identified with a raw spatial reflection amplitude without matching the
channel normalizations.

\subsection{Background energy and state dependence}

If $|\beta|^2$ is nonzero, energy must come from the background or the device
maintaining it. Rotating and charged black holes are examples of such energy
reservoirs. The construction of the positive-frequency bases, horizon data,
and quantum state decides whether a given mode contributes to emission.
The charged-black-hole analysis in Ref.~\cite{AlvarezWinstanley2025} supplies
an explicit example of this state dependence; it is cited here as prior work,
not as a new consequence of the present one-dimensional model. The general
calculation above does not determine the numerical $\beta$ coefficients for
an arbitrary black hole.

\section{From fermionic constituents to a pair-condensate effective field}
\label{sec:pairs}

\subsection{Two-body coordinates and pair mass}

\paragraph{Step 1: separate center-of-mass and relative coordinates.}
For two equal-mass fermions in opposite spin states, let
\begin{equation}
 \bm R=\frac{\bm r_1+\bm r_2}{2},\qquad
 \bm\rho=\bm r_1-\bm r_2.
\end{equation}
The chain rule yields
\begin{equation}
 \nabla_1=\tfrac12\nabla_R+\nabla_\rho,\qquad
 \nabla_2=\tfrac12\nabla_R-\nabla_\rho,
\end{equation}
and therefore
\begin{equation}
 \nabla_1^2+\nabla_2^2=
 \tfrac12\nabla_R^2+2\nabla_\rho^2.
\end{equation}
For an interaction $U_{\mathrm{int}}(\rho)$, the two-body Hamiltonian becomes
\begin{equation}
 H_2=-\frac{\hbar^2}{4m_{\mathrm f}}\nabla_R^2
      -\frac{\hbar^2}{m_{\mathrm f}}\nabla_\rho^2
      +U_{\mathrm{int}}(\rho).
 \label{eq:two-body-hamiltonian}
\end{equation}
The first term is a center-of-mass kinetic term with mass
$M_{\mathrm B}=2m_{\mathrm f}$.

\paragraph{Step 2: specify a bound internal state.}
Suppose
\begin{equation}
 \left[-\frac{\hbar^2}{m_{\mathrm f}}\nabla_\rho^2
 +U_{\mathrm{int}}(\rho)\right]\varphi(\bm\rho)
 =-E_{\mathrm b}\varphi(\bm\rho),\qquad E_{\mathrm b}>0,
\end{equation}
with $\int|\varphi|^2\dd^3\rho=1$. A factorized state is
\begin{equation}
 \psi(\bm r_1,\bm r_2)=
 \ee^{\ii\bm K\cdot\bm R}\varphi(\bm\rho),\qquad
 E=\frac{\hbar^2K^2}{2M_{\mathrm B}}-E_{\mathrm b}.
 \label{eq:pair-state}
\end{equation}
For a spin-singlet pair the spin state is antisymmetric and the spatial state
can be symmetric under $\bm\rho\mapsto-\bm\rho$. The complete fermionic
state remains antisymmetric. In the isotropic Galilean two-body model
\eqref{eq:two-body-hamiltonian}, finite $\bm K$ does not by itself break the
relative rotational symmetry: center-of-mass and internal motions remain
separable. A medium, a lattice, or spin-dependent interactions would need to
be specified to change that statement.

\paragraph{Step 3: display the momentum-pair content.}
Using the Fourier expansion
$\varphi(\bm\rho)=\sum_{\bm k}\varphi_{\bm k}
\ee^{\ii\bm k\cdot\bm\rho}$ in a normalization box, we obtain
\begin{equation}
 \psi=\sum_{\bm k}\varphi_{\bm k}
 \ee^{\ii(\bm K/2+\bm k)\cdot\bm r_1}
 \ee^{\ii(\bm K/2-\bm k)\cdot\bm r_2}.
\end{equation}
Each term contains two constituent momenta whose sum is $\hbar\bm K$.
This rewriting specifies a pair state; it does not yet establish a
macroscopic condensate or a scattering gain.

\subsection{Why a pair is only approximately an elementary boson}

For simplicity set $\bm K=0$ and define
\begin{equation}
 \hat B^\dagger=\sum_{\bm k}\varphi_{\bm k}
 \hat c_{\bm k\uparrow}^\dagger\hat c_{-\bm k\downarrow}^\dagger,
 \qquad\sum_{\bm k}|\varphi_{\bm k}|^2=1.
\end{equation}
Write $\hat d_{\bm k}=\hat c_{-\bm k\downarrow}\hat c_{\bm k\uparrow}$.
For a fixed $\bm k$, anticommuting the constituent operators gives
\begin{align}
 \hat d_{\bm k}\hat d_{\bm k}^\dagger
 &=(1-\hat n_{\bm k\uparrow})(1-\hat n_{-\bm k\downarrow}),\\
 \hat d_{\bm k}^\dagger\hat d_{\bm k}
 &=\hat n_{\bm k\uparrow}\hat n_{-\bm k\downarrow}.
\end{align}
Subtracting cancels the product of occupation operators. Distinct pair modes
use distinct constituent modes and their even bilinears commute. Consequently
\begin{equation}
 [\hat B,\hat B^\dagger]=1-
 \sum_{\bm k}|\varphi_{\bm k}|^2
 (\hat n_{\bm k\uparrow}+\hat n_{-\bm k\downarrow}).
 \label{eq:pair-commutator}
\end{equation}
The correction is small only when constituent-state overlap is small in the
relevant many-body state. A dilute gas of well-separated molecules is the
regime used here. The effective boson field does not remove the Pauli principle
from the microscopic Hilbert space.

\subsection{Effective energy and the Gross--Pitaevskii equation}

We now assume that the low-energy pair degree of freedom can be represented
by a neutral complex field $\Psi_{\mathrm B}$, normalized so that
$n_{\mathrm B}=|\Psi_{\mathrm B}|^2$ is the pair number density. The
strong-coupling reduction from paired fermions is an established microscopic
route \cite{PieriStrinati2003}; here we use its leading effective structure,
not a new derivation of its coefficients. In three dimensions take
\begin{equation}
 E_{\mathrm B}[\Psi_{\mathrm B}]=\int\dd^3x\left[
 \frac{\hbar^2}{2M_{\mathrm B}}|\nabla\Psi_{\mathrm B}|^2
 +U_{\mathrm B}|\Psi_{\mathrm B}|^2
 +\frac{g_{\mathrm B}}{2}|\Psi_{\mathrm B}|^4\right],
 \qquad g_{\mathrm B}=\frac{4\pi\hbar^2a_{\mathrm B}}{M_{\mathrm B}}>0.
 \label{eq:pair-energy}
\end{equation}
The dimer-dimer scattering length $a_{\mathrm B}$ is an effective parameter
determined from the microscopic problem, not merely by assigning bosonic
statistics \cite{Petrov2004}. This energy has the Gross--Pitaevskii form
\cite{Gross1961,Pitaevskii1961}.

\paragraph{Step 1: vary the gradient term.}
For variations vanishing on the integration boundary, integration by parts
gives
\begin{equation}
 \delta_{\Psi_{\mathrm B}^*}
 \int\frac{\hbar^2}{2M_{\mathrm B}}\nabla\Psi_{\mathrm B}^*\cdot
 \nabla\Psi_{\mathrm B}\,\dd^3x
 =-\int\frac{\hbar^2}{2M_{\mathrm B}}\nabla^2\Psi_{\mathrm B}
 \delta\Psi_{\mathrm B}^*\,\dd^3x.
\end{equation}
For the local terms,
\begin{equation}
 \frac{\delta E_{\mathrm B}}{\delta\Psi_{\mathrm B}^*}
 =-\frac{\hbar^2}{2M_{\mathrm B}}\nabla^2\Psi_{\mathrm B}
 +U_{\mathrm B}\Psi_{\mathrm B}
 +g_{\mathrm B}|\Psi_{\mathrm B}|^2\Psi_{\mathrm B}.
\end{equation}

\paragraph{Step 2: include the time-derivative term.}
Use the action
\begin{equation}
 S_{\mathrm B}=\int\dd t\left\{\int\dd^3x\,
 \frac{\ii\hbar}{2}
 (\Psi_{\mathrm B}^*\partial_t\Psi_{\mathrm B}
 -\Psi_{\mathrm B}\partial_t\Psi_{\mathrm B}^*)-E_{\mathrm B}\right\}.
\end{equation}
Its Euler--Lagrange equation with respect to $\Psi_{\mathrm B}^*$ yields
\begin{equation}
 \boxed{\ii\hbar\partial_t\Psi_{\mathrm B}=
 \left[-\frac{\hbar^2}{2M_{\mathrm B}}\nabla^2+U_{\mathrm B}
 +g_{\mathrm B}|\Psi_{\mathrm B}|^2\right]\Psi_{\mathrm B}.}
 \label{eq:gp}
\end{equation}
Equation~\eqref{eq:gp} is an effective equation for a pair condensate. It is
neither the one-particle Dirac equation nor a conclusion obtainable from a
boundary condition on that equation alone.

The assumed regime includes $n_{\mathrm B}a_{\mathrm B}^3\ll1$,
$g_{\mathrm B}n_{\mathrm B}\ll E_{\mathrm b}$, excitation energies small
compared with $E_{\mathrm b}$, and background variation lengths large compared
with the pair size. Charged pairs would require electromagnetic dynamics;
the neutral model avoids equating charged-superconductor modes with neutral
sound modes.

\section{Collective excitations and their conserved norm}
\label{sec:bdg}

\subsection{Linearization around a stationary condensate}

Let
\begin{equation}
 \Psi_{\mathrm B}(\bm x,t)=\ee^{-\ii\mu_{\mathrm B}t/\hbar}
 [\Psi_0(\bm x)+\delta\psi(\bm x,t)],
\end{equation}
where
\begin{equation}
 [H_0+g_{\mathrm B}|\Psi_0|^2]\Psi_0=\mu_{\mathrm B}\Psi_0,
 \qquad H_0=-\frac{\hbar^2}{2M_{\mathrm B}}\nabla^2+U_{\mathrm B}.
\end{equation}
Expanding the nonlinear term through first order gives
\begin{align}
 |\Psi_0+\delta\psi|^2(\Psi_0+\delta\psi)
 &=|\Psi_0|^2\Psi_0
 +2|\Psi_0|^2\delta\psi+\Psi_0^2\delta\psi^*
 +O(\delta\psi^2).
\end{align}
Subtracting the background equation from Eq.~\eqref{eq:gp},
\begin{equation}
 \ii\hbar\partial_t\delta\psi=L\delta\psi+\Delta_{\mathrm B}\delta\psi^*,
 \quad L=H_0-\mu_{\mathrm B}+2g_{\mathrm B}|\Psi_0|^2,
 \quad\Delta_{\mathrm B}=g_{\mathrm B}\Psi_0^2.
\end{equation}
Substitute
$\delta\psi=u\ee^{-\ii\omega t}+v^*\ee^{\ii\omega t}$ and equate the
coefficients of each time dependence. This yields
\begin{equation}
 \hbar\omega\begin{pmatrix}u\\v\end{pmatrix}
 =\mathcal B\begin{pmatrix}u\\v\end{pmatrix},\qquad
 \mathcal B=\begin{pmatrix}L&\Delta_{\mathrm B}\\
 -\Delta_{\mathrm B}^*&-L^*\end{pmatrix}.
 \label{eq:bosonic-bdg}
\end{equation}

\subsection{The indefinite collective norm}

For boundary conditions on which $L$ is self-adjoint,
\begin{equation}
 \mathcal B^\dagger\sigma_3=\sigma_3\mathcal B,
 \qquad\sigma_3=\diag(1,-1).
\end{equation}
For a wave packet $w=(u,v)^{\mathsf T}$ satisfying
$\ii\hbar\partial_t w=\mathcal B w$, differentiate
\begin{align}
 \frac{\dd}{\dd t}\int w^\dagger\sigma_3w\,\dd^3x
 &=\frac{\ii}{\hbar}\int w^\dagger
 (\mathcal B^\dagger\sigma_3-\sigma_3\mathcal B)w\,\dd^3x=0.
\end{align}
Thus the conserved collective norm is
\begin{equation}
 (w,w)_{\mathrm B}=\int(|u|^2-|v|^2)\,\dd^3x.
 \label{eq:bdg-norm}
\end{equation}
It is indefinite. The familiar fermionic quasiparticle BdG matrix, in a
two-component spin-singlet convention, instead has the Hermitian structure
\begin{equation}
 \mathcal H_{\mathrm F}=\begin{pmatrix}h&\Delta_{\mathrm F}\\
 \Delta_{\mathrm F}^*&-h^*\end{pmatrix},
\end{equation}
with a positive quasiparticle normalization. Its components must not be
assigned the norm in Eq.~\eqref{eq:bdg-norm}. Bosonic condensate fluctuations
and fermionic quasiparticles are different excitations, even when both are
described using the name Bogoliubov--de Gennes.

\subsection{Uniform-flow dispersion and the long-wavelength limit}

In a uniform condensate at rest with $\Psi_0=\sqrt{n_0}$ and
$\mu_{\mathrm B}=U_{\mathrm B}+g_{\mathrm B}n_0$, a Fourier component has
\begin{equation}
 \mathcal B_{\bm k}=
 \begin{pmatrix}\varepsilon_k+g_{\mathrm B}n_0&g_{\mathrm B}n_0\\
 -g_{\mathrm B}n_0&-\varepsilon_k-g_{\mathrm B}n_0\end{pmatrix},
 \qquad\varepsilon_k=\frac{\hbar^2k^2}{2M_{\mathrm B}}.
\end{equation}
The determinant condition is
\begin{align}
 0&=(\varepsilon_k+g_{\mathrm B}n_0-\hbar\Omega)
 (-\varepsilon_k-g_{\mathrm B}n_0-\hbar\Omega)
 +(g_{\mathrm B}n_0)^2\nonumber\\
 &=(\hbar\Omega)^2-\varepsilon_k(\varepsilon_k+2g_{\mathrm B}n_0).
\end{align}
A Galilean boost to uniform velocity $\bm v_0$ gives
$\Omega=\omega-\bm v_0\cdot\bm k$, hence
\begin{equation}
 \boxed{(\omega-\bm v_0\cdot\bm k)^2=
 \cs^2 k^2+\frac{\hbar^2k^4}{4M_{\mathrm B}^2},\qquad
 \cs^2=\frac{g_{\mathrm B}n_0}{M_{\mathrm B}}.}
 \label{eq:bogoliubov-dispersion}
\end{equation}
The corresponding healing length is
\begin{equation}
 \xi=\frac{\hbar}{\sqrt{2M_{\mathrm B}g_{\mathrm B}n_0}}
 =\frac{\hbar}{\sqrt2M_{\mathrm B}\cs}.
\end{equation}
The quartic term is negligible relative to $\cs^2k^2$ when $k\xi\ll1$.
For the positive-energy rest-frame branch, the eigenvector has positive
norm; its negative-frequency partner has negative norm. A moving background
can make a positive laboratory frequency coexist with negative comoving
frequency and negative collective norm. This is the mode conversion relevant
to collective amplification.

\section{Hydrodynamic derivation of the collective wave equation}
\label{sec:hydro}

\subsection{Amplitude-phase decomposition}

Write $\Psi_{\mathrm B}=\sqrt n\,\ee^{\ii\theta}$ and define
\begin{equation}
 \bm v=\frac{\hbar}{M_{\mathrm B}}\nabla\theta.
\end{equation}
Away from phase singularities the flow is irrotational. The derivatives are
\begin{align}
 \partial_t\Psi_{\mathrm B}
 &=\ee^{\ii\theta}\left(\frac{\partial_t n}{2\sqrt n}
 +\ii\sqrt n\,\partial_t\theta\right),\\
 \nabla^2\Psi_{\mathrm B}
 &=\ee^{\ii\theta}\left[\nabla^2\sqrt n-\sqrt n(\nabla\theta)^2
 +\ii\left(2\nabla\sqrt n\cdot\nabla\theta
 +\sqrt n\,\nabla^2\theta\right)\right].
\end{align}
Substitute these into Eq.~\eqref{eq:gp}. Equating imaginary parts and
multiplying by $2\sqrt n/\hbar$ gives
\begin{equation}
 \partial_t n+\frac{\hbar}{M_{\mathrm B}}
 [\nabla n\cdot\nabla\theta+n\nabla^2\theta]=0,
\end{equation}
which is
\begin{equation}
 \partial_t n+\nabla\cdot(n\bm v)=0.
 \label{eq:pair-continuity}
\end{equation}
Equating real parts and dividing by $\sqrt n$ yields
\begin{equation}
 \hbar\partial_t\theta+\frac{M_{\mathrm B}}2v^2+U_{\mathrm B}
 +g_{\mathrm B}n-\frac{\hbar^2}{2M_{\mathrm B}}
 \frac{\nabla^2\sqrt n}{\sqrt n}=0.
 \label{eq:bernoulli}
\end{equation}
The last term is the quantum-pressure contribution.

\subsection{Linearized density and phase}

Let $n=n_0+\delta n$ and $\theta=\theta_0+\vartheta$. The background is
stationary, although $\theta_0$ includes $-\mu_{\mathrm B}t/\hbar$.
Define $D_t=\partial_t+\bm v_0\cdot\nabla$ and
$\delta\bm v=(\hbar/M_{\mathrm B})\nabla\vartheta$.
Neglecting the linearized quantum pressure in the long-wavelength regime,
the first-order part of Eq.~\eqref{eq:bernoulli} is
\begin{equation}
 \hbar\partial_t\vartheta+M_{\mathrm B}\bm v_0\cdot\delta\bm v
 +g_{\mathrm B}\delta n=0.
\end{equation}
Substituting $\delta\bm v$ gives
\begin{equation}
 \boxed{\delta n=-\frac{\hbar}{g_{\mathrm B}}D_t\vartheta.}
 \label{eq:density-phase}
\end{equation}
The linearized continuity equation is
\begin{equation}
 \partial_t\delta n+
 \nabla\cdot\left(\bm v_0\delta n+
 \frac{\hbar n_0}{M_{\mathrm B}}\nabla\vartheta\right)=0.
\end{equation}
Inserting Eq.~\eqref{eq:density-phase} and dividing by $-\hbar$ gives
\begin{equation}
 \boxed{\partial_t\left(\frac1{g_{\mathrm B}}D_t\vartheta\right)
 +\nabla\cdot\left[\frac{\bm v_0}{g_{\mathrm B}}D_t\vartheta
 -\frac{n_0}{M_{\mathrm B}}\nabla\vartheta\right]=0.}
 \label{eq:acoustic-pde}
\end{equation}
The coefficient $1/g_{\mathrm B}$ is retained inside derivatives so that its
origin is transparent; the explicit example below takes it to be constant.

\subsection{Action, acoustic geometry, and the norm sign}

Equation~\eqref{eq:acoustic-pde} follows from the real quadratic action
\begin{equation}
 S^{(2)}=\frac{\hbar^2}{2}\int\dd t\,\dd^3x
 \left[\frac1{g_{\mathrm B}}(D_t\vartheta)^2
 -\frac{n_0}{M_{\mathrm B}}(\nabla\vartheta)^2\right].
 \label{eq:acoustic-action}
\end{equation}
Indeed, varying $\vartheta$ and integrating its time and spatial derivatives
by parts gives the negative of the left side of
Eq.~\eqref{eq:acoustic-pde}. In three spatial dimensions, define the acoustic
metric in signature $(+,-,-,-)$ by
\begin{equation}
 \dd s_{\mathrm{ac}}^2=\frac{n_0}{M_{\mathrm B}\cs}
 \left[(\cs^2-v_0^2)\dd t^2
 +2\bm v_0\cdot\dd\bm x\,\dd t-\dd\bm x^2\right].
 \label{eq:acoustic-metric}
\end{equation}
Its densitized inverse is
\begin{equation}
 \sqrt{|g_{\mathrm{ac}}|}\,g_{\mathrm{ac}}^{\mu\nu}
 =\frac1{g_{\mathrm B}}
 \begin{pmatrix}1&\bm v_0^{\mathsf T}\\
 \bm v_0&\bm v_0\bm v_0^{\mathsf T}-\cs^2\mathbf1\end{pmatrix},
\end{equation}
since $\cs^2=g_{\mathrm B}n_0/M_{\mathrm B}$. Thus
Eq.~\eqref{eq:acoustic-pde} is a scalar wave equation in the effective
geometry. This is the analogue-gravity construction
\cite{Unruh1981,Visser1998}; the metric is a propagation metric for sound,
not a solution of Einstein's equation sourced by the condensate.

Complexifying linear solutions, a conserved inner product is, up to a positive
overall normalization,
\begin{equation}
 (\vartheta_1,\vartheta_2)_{\mathrm{ac}}
 =\ii\hbar^2\int\frac{\dd^3x}{g_{\mathrm B}}
 [\vartheta_1^*D_t\vartheta_2
 -(D_t\vartheta_1)^*\vartheta_2].
 \label{eq:acoustic-norm}
\end{equation}
For a uniform-flow plane wave,
$D_t\vartheta=-\ii(\omega-\bm v_0\cdot\bm k)\vartheta$, so its norm density
is
\begin{equation}
 \frac{2\hbar^2}{g_{\mathrm B}}
 (\omega-\bm v_0\cdot\bm k)|\vartheta|^2.
\end{equation}
The sign is that of the comoving frequency. The positive definiteness of the
constituent Dirac number density places no positivity requirement on this
different collective-wave norm.

\section{A conditional rotating-flow model and its amplification identity}
\label{sec:rotating}

\subsection{A stationary background satisfying the effective equations}

Consider an axially invariant condensate, with perturbations independent of
$z$, outside a draining core. Take $n_0$, $g_{\mathrm B}$, and $\cs$ constant
and choose
\begin{equation}
 v_r=-\frac{D}{r},\qquad v_\phi=\frac{C}{r},\qquad
 D>0,
 \label{eq:draining-flow}
\end{equation}
where $D$ and $C$ have dimensions length squared per time. The continuity
equation is satisfied for $r>0$ because
\begin{equation}
 \nabla\cdot(n_0\bm v_0)=\frac1r\partial_r(rn_0v_r)
 +\frac1r\partial_\phi(n_0v_\phi)=0.
\end{equation}
A background phase producing this flow is
\begin{equation}
 \theta_0=-\frac{\mu_{\mathrm B}t}{\hbar}
 -\frac{M_{\mathrm B}D}{\hbar}\ln\frac r{r_0}
 +\frac{M_{\mathrm B}C}{\hbar}\phi.
\end{equation}
Single-valuedness of the condensate requires
\begin{equation}
 \oint\bm v_0\cdot\dd\bm\ell=2\pi C
 =\frac{2\pi\hbar\ell}{M_{\mathrm B}},\qquad
 C=\frac{\hbar\ell}{M_{\mathrm B}},\quad\ell\in\mathbb Z.
 \label{eq:circulation}
\end{equation}
Here $\ell$ labels the background circulation and is distinct from the
perturbation's azimuthal number $m$.

Since $n_0$ is constant, the background quantum pressure vanishes. Substituting
the phase into Eq.~\eqref{eq:bernoulli} fixes the required exterior potential:
\begin{equation}
 U_{\mathrm B}(r)=\mu_{\mathrm B}-g_{\mathrm B}n_0
 -\frac{M_{\mathrm B}(D^2+C^2)}{2r^2}.
 \label{eq:supporting-potential}
\end{equation}
Thus the proposed flow is consistent with the exterior effective background
equations when this potential is supplied. Its continuation to $r=0$ is not
part of the model. A sink inside the exterior domain and an outer reservoir
are needed to maintain the inward pair flux. Establishing their microscopic
implementation and the stability of a multiply wound flow is a separate
problem.

\subsection{Horizon, ergosurface, and corotating coordinates}

The normal flow reaches the sound speed at
\begin{equation}
 \rH=\frac D{\cs},\qquad
 \OH=\frac{v_\phi(\rH)}{\rH}=\frac C{\rH^2}.
 \label{eq:horizon-omega}
\end{equation}
The stationary time direction becomes null when $v_r^2+v_\phi^2=\cs^2$,
so the ergosurface is at
\begin{equation}
 r_{\mathrm e}=\frac{\sqrt{D^2+C^2}}{\cs}\geq\rH.
\end{equation}
Dropping the constant positive conformal factor in
Eq.~\eqref{eq:acoustic-metric}, the $z$-independent line element is
\begin{equation}
 \dd s^2=\left[\cs^2-\frac{D^2+C^2}{r^2}\right]\dd t^2
 -\frac{2D}{r}\dd t\,\dd r+2C\dd t\,\dd\phi
 -\dd r^2-r^2\dd\phi^2.
 \label{eq:acoustic-polar}
\end{equation}
Introduce $F=1-\rH^2/r^2$ and the exterior coordinate differentials
\begin{equation}
 \dd t=\dd\tau+\frac{D}{r\cs^2F}\dd r,\qquad
 \dd\phi=\dd\varphi+\frac{CD}{r^3\cs^2F}\dd r.
 \label{eq:bl-change}
\end{equation}
The $\dd\tau\dd r$ coefficient vanishes because
\begin{equation}
 \left(\cs^2-\frac{D^2+C^2}{r^2}\right)
 \frac{D}{r\cs^2F}+C\frac{CD}{r^3\cs^2F}-\frac Dr=0.
\end{equation}
The $\dd\varphi\dd r$ coefficient vanishes because
$CD/(r\cs^2F)-r^2CD/(r^3\cs^2F)=0$. Substitution gives
\begin{equation}
 \dd s^2=\left(\cs^2F-\frac{C^2}{r^2}\right)\dd\tau^2
 +2C\dd\tau\dd\varphi-\frac{\dd r^2}{F}-r^2\dd\varphi^2.
 \label{eq:bl-metric}
\end{equation}
These exterior coordinates are singular at the horizon; ingoing regularity is
defined using the original flow coordinates. The generator
$\partial_\tau+\OH\partial_\varphi$ acts on a mode
$\ee^{-\ii\omega\tau+\ii m\varphi}$ with frequency
$\omega-m\OH$.

\subsection{Deriving the radial equation}

\paragraph{Step 1: compute the inverse metric.}
The determinant of the $(\tau,\varphi)$ block is
$-\cs^2Fr^2$, and the determinant of the full three-coordinate metric is
$\cs^2r^2$. Therefore
\begin{equation}
 \sqrt{|g|}=\cs r,\qquad
 g^{\tau\tau}=\frac1{\cs^2F},\quad
 g^{\tau\varphi}=\frac C{\cs^2Fr^2},\quad
 g^{\varphi\varphi}=-\frac1{r^2}+\frac{C^2}{\cs^2Fr^4},\quad
 g^{rr}=-F.
\end{equation}
This reduction is valid for the constant conformal factor and $z$-independent
background being used; a variable density requires retaining the appropriate
densitized coefficients.

\paragraph{Step 2: separate the mode.}
For
\begin{equation}
 \vartheta=\ee^{-\ii\omega\tau+\ii m\varphi}\mathcal U(r),
 \qquad m\in\mathbb Z,
\end{equation}
the time and angular terms of the wave equation combine to
\begin{equation}
 \left[-\frac{\omega^2}{\cs^2F}
 +\frac{2\omega mC}{\cs^2Fr^2}
 -\frac{m^2C^2}{\cs^2Fr^4}+\frac{m^2}{r^2}\right]\mathcal U
 =\left[-\frac{(\omega-mC/r^2)^2}{\cs^2F}
 +\frac{m^2}{r^2}\right]\mathcal U.
\end{equation}
The radial term is $-r^{-1}\partial_r(rF\partial_r\mathcal U)$.
Multiplying the separated equation by $-1$ gives
\begin{equation}
 \frac1r\frac{\dd}{\dd r}\left(rF\frac{\dd\mathcal U}{\dd r}\right)
 +\left[\frac{(\omega-mC/r^2)^2}{\cs^2F}-\frac{m^2}{r^2}\right]
 \mathcal U=0.
 \label{eq:acoustic-radial-raw}
\end{equation}

\paragraph{Step 3: introduce a tortoise coordinate.}
Define $\dd r_*/\dd r=1/F$. Integration yields
\begin{equation}
 r_*=r+\frac{\rH}{2}\ln\left|\frac{r-\rH}{r+\rH}\right|
 +\text{constant}.
 \label{eq:tortoise}
\end{equation}
Multiplying Eq.~\eqref{eq:acoustic-radial-raw} by $F$ and denoting
$\dd/\dd r_*$ by a dot for this calculation gives
\begin{equation}
 \ddot{\mathcal U}+\frac Fr\dot{\mathcal U}
 +\left[\frac{(\omega-mC/r^2)^2}{\cs^2}-\frac{Fm^2}{r^2}\right]
 \mathcal U=0.
\end{equation}

\paragraph{Step 4: remove the first derivative.}
Set $\mathcal U=G/\sqrt r$ and $P=F/r$. Since
$\dot r=F$, differentiation gives
\begin{align}
 \dot{\mathcal U}&=r^{-1/2}(\dot G-\tfrac12PG),\\
 \ddot{\mathcal U}&=r^{-1/2}
 [\ddot G-P\dot G+(\tfrac14P^2-\tfrac12\dot P)G].
\end{align}
Inserting these expressions cancels the $\dot G$ terms. Moreover
\begin{equation}
 \dot P=F\left(\frac{F'}r-\frac F{r^2}\right),
 \qquad F'=\frac{\dd F}{\dd r}.
\end{equation}
The resulting equation is
\begin{equation}
 \boxed{\frac{\dd^2G}{\dd r_*^2}
 +\left[\frac{(\omega-mC/r^2)^2}{\cs^2}-V_m(r)\right]G=0,}
 \label{eq:acoustic-radial}
\end{equation}
with real effective potential
\begin{equation}
 V_m(r)=F\left[\frac{m^2}{r^2}+\frac{F'}{2r}
 -\frac F{4r^2}\right].
 \label{eq:acoustic-potential}
\end{equation}
All terms in the square bracket of Eq.~\eqref{eq:acoustic-radial} have
dimensions length to the power $-2$. This equation is the explicit
collective-wave scattering problem, with the same draining-vortex mechanism
studied in Ref.~\cite{BasakMajumdar2003}.

\subsection{Flux conservation and the superradiant interval}

For real $\omega$, the radial coefficient in Eq.~\eqref{eq:acoustic-radial}
is real. Therefore
\begin{equation}
 W=G^*\frac{\dd G}{\dd r_*}
 -G\frac{\dd G^*}{\dd r_*},\qquad
 \frac{\dd W}{\dd r_*}=0.
\end{equation}
The asymptotic boundary conditions are
\begin{align}
 G&\sim\mathcal I\ee^{-\ii\omega r_*/\cs}
 +\mathcal R\ee^{\ii\omega r_*/\cs},
 &&r_*\longrightarrow+\infty,\\
 G&\sim\mathcal T\ee^{-\ii(\omega-m\OH)r_*/\cs},
 &&r_*\longrightarrow-\infty.
 \label{eq:acoustic-boundaries}
\end{align}
The second condition is the horizon-ingoing branch, including when
$\omega-m\OH<0$. Computing the Wronskian on each side gives
\begin{equation}
 W_\infty=\frac{2\ii\omega}{\cs}
 (|\mathcal R|^2-|\mathcal I|^2),\qquad
 W_{\mathrm H}=-\frac{2\ii(\omega-m\OH)}{\cs}|\mathcal T|^2.
\end{equation}
Equating them and dividing by $2\ii\omega|\mathcal I|^2/\cs$ yields
\begin{equation}
 \boxed{\Refl_{\mathrm{coll}}=
 \frac{|\mathcal R|^2}{|\mathcal I|^2}
 =1-\frac{\omega-m\OH}{\omega}
 \frac{|\mathcal T|^2}{|\mathcal I|^2}.}
 \label{eq:collective-gain}
\end{equation}
For $\omega>0$ and $m\OH>0$, nonzero transmission therefore gives
\begin{equation}
 \boxed{0<\omega<m\OH\quad\Longrightarrow\quad
 \Refl_{\mathrm{coll}}>1.}
 \label{eq:collective-window}
\end{equation}
At the asymptotic boundary the background flow tends to zero, so the incident
and reflected positive-frequency modes have identical energy-flux
normalizations. Their amplitude-squared ratio is consequently an energy-flux
ratio. The negative absorbed collective norm in
Eq.~\eqref{eq:collective-gain} corresponds to negative laboratory Killing
energy entering the acoustic horizon. The reflected excess is drawn from the
moving background; a finite unmaintained flow loses energy and angular
momentum. This does not require negative constituent Dirac number density.

\subsection{Necessary conditions and limiting checks}

If $C=0$, then $\OH=0$ and Eq.~\eqref{eq:collective-gain} gives ordinary
attenuation for every $\omega>0$. The same is true for $m=0$. At the threshold
$\omega=m\OH$, the ideal horizon norm flux vanishes and the limiting
reflection ratio is unity for a regular scattering solution. If
$\mathcal T=0$, there is no gain even inside the frequency interval.
Thus the interval identifies an available negative-norm channel, while the
radial dynamics determines how strongly it couples.

The required approximations are not automatic. One needs $\xi\ll\rH$ and
long-wavelength propagation in the exterior scattering region, intact pairs,
and an inner condition that absorbs the relevant mode. Dispersion can alter
near-core mode conversion and boundary data. A perfectly reflecting regular
core, additional channels, or a dynamically unstable background requires
solving a different problem. The microscopic scattering calculation for a
specific fermionic superfluid is not supplied by the nondispersive acoustic
equation alone.

\section{A reproducible numerical example in the acoustic model}
\label{sec:numerics}

\subsection{Dimensionless parameters and initial data}

Scale lengths by $\rH$, velocities by $\cs$, and frequencies by
$\cs/\rH$. Set
\begin{equation}
 \rH=\cs=D=1,\qquad C=\tfrac12,\qquad m=1,
 \qquad\OH=\tfrac12.
 \label{eq:numerical-parameters}
\end{equation}
These are dimensionless acoustic parameters. They do not fix the absolute
constituent mass or density. Circulation quantization must still be respected
in a pair-condensate realization. In particular,
\begin{equation}
 \frac C{\cs\rH}=\sqrt2\,\ell\frac\xi{\rH}.
 \label{eq:circulation-scale}
\end{equation}
The choice $C/(\cs\rH)=1/2$ is compatible, for example, with
$\ell=40$ and $\xi/\rH=1/(80\sqrt2)\simeq0.00884$ at the level of scale
separation. This does not prove the stability or preparation of that multiply
wound draining background. The numerical example tests the effective acoustic
equation, not that preparation problem.

Let $x=r-1$ and $k_{\mathrm H}=\omega-C$ for these parameters. Near the
horizon,
\begin{equation}
 F=2x-3x^2+O(x^3),\qquad
 \frac{F'}F=\frac1x-\frac32+O(x).
\end{equation}
In the ordinary $r$ coordinate, Eq.~\eqref{eq:acoustic-radial} reads
\begin{equation}
 G''+\frac{F'}F G'
 +\left[\frac{(\omega-C/r^2)^2}{F^2}-\frac{V_1}{F^2}\right]G=0.
\end{equation}
The coefficient of $G$ has the expansion
\begin{equation}
 \frac{k_{\mathrm H}^2}{4x^2}+
 \frac{q_{-1}}x+O(1),\qquad
 q_{-1}=\frac34k_{\mathrm H}^2+Ck_{\mathrm H}-1.
\end{equation}
Insert $G=x^p(1+a_1x+O(x^2))$. The leading coefficient gives
\begin{equation}
 p(p-1)+p+\frac{k_{\mathrm H}^2}{4}=0,
 \qquad p=-\frac{\ii k_{\mathrm H}}2
\end{equation}
for the ingoing branch. At the next order,
\begin{equation}
 [(p+1)^2+k_{\mathrm H}^2/4]a_1-\tfrac32p+q_{-1}=0.
\end{equation}
Using $p^2=-k_{\mathrm H}^2/4$ gives the explicit correction
\begin{equation}
 a_1=-\frac{-\tfrac32p+q_{-1}}{1+2p}.
 \label{eq:frobenius-coefficient}
\end{equation}
At $r=1+\epsilon$, we initialize
\begin{equation}
 G=\epsilon^p(1+a_1\epsilon),\qquad
 P=F\frac{\dd G}{\dd r}
 =F\epsilon^{p-1}[p+(p+1)a_1\epsilon].
\end{equation}
The integration system is
\begin{equation}
 \frac{\dd G}{\dd r}=\frac PF,\qquad
 \frac{\dd P}{\dd r}=-\frac{(\omega-C/r^2)^2-V_1}{F}G.
 \label{eq:numerical-system}
\end{equation}
We use the DOP853 implementation in SciPy, relative tolerance
$2\times10^{-11}$, absolute tolerance $2\times10^{-13}$,
$\epsilon=10^{-6}$, and $r_{\max}=400$.

\subsection{Amplitude extraction and flux check}

At the outer boundary, temporarily neglecting the decaying potential tail,
$G=\mathcal I\ee^{-\ii\omega r_*}+\mathcal R\ee^{\ii\omega r_*}$ and
$P=-\ii\omega\mathcal I\ee^{-\ii\omega r_*}
+\ii\omega\mathcal R\ee^{\ii\omega r_*}$. Adding and subtracting these
equations gives
\begin{equation}
 \mathcal I=\tfrac12(G+\ii P/\omega)\ee^{\ii\omega r_*},\qquad
 \mathcal R=\tfrac12(G-\ii P/\omega)\ee^{-\ii\omega r_*}.
 \label{eq:amplitude-extraction}
\end{equation}
The Frobenius initialization sets the magnitude of the horizon amplitude to
unity to the retained order. We calculate
\begin{equation}
 \Refl_{\mathrm{coll}}=|\mathcal R/\mathcal I|^2,\qquad
 T_{\mathrm{amp}}=|\mathcal T/\mathcal I|^2=1/|\mathcal I|^2,
\end{equation}
and monitor the residual
\begin{equation}
 \delta_{\mathrm{flux}}=
 \left|\Refl_{\mathrm{coll}}+
 \frac{\omega-1/2}{\omega}T_{\mathrm{amp}}-1\right|.
\end{equation}
The results are shown in Table~\ref{tab:numerics}.

\begin{table}[htbp]
\centering
\caption{Numerical reflection in the nondispersive acoustic model
\eqref{eq:numerical-parameters}. The amplitudes are normalized as in
Eq.~\eqref{eq:amplitude-extraction}; $T_{\mathrm{amp}}$ is not a positive
transmitted energy fraction. Entries are finite-radius estimates.}
\label{tab:numerics}
\begin{tabular}{rrrr}
\toprule
$\omega$ & $\Refl_{\mathrm{coll}}$ & $T_{\mathrm{amp}}$ & $100 Z$ (percent)\\
\midrule
0.15 & 1.02224 & 0.00953 &  2.224\\
0.25 & 1.04214 & 0.04214 &  4.214\\
0.35 & 1.05058 & 0.11801 &  5.058\\
0.45 & 1.02946 & 0.26512 &  2.946\\
0.50 & 1.00000 & 0.37483 &  0.000\\
0.60 & 0.88777 & 0.67339 & -11.223\\
0.80 & 0.48448 & 1.37473 & -51.552\\
\bottomrule
\end{tabular}
\end{table}

Across the seven listed frequencies, the largest flux residual is below
$7.3\times10^{-11}$. Changing $r_{\max}$ from $200$ to $400$ changes the
reflection ratio by at most $6.3\times10^{-6}$. Changing the starting offset
from $10^{-6}$ to $3\times10^{-6}$ while tightening the relative tolerance to
$5\times10^{-12}$ changes it by less than $3.4\times10^{-12}$.
The first comparison estimates sensitivity to the outer extraction radius;
it is not a certified bound on the remaining infinite-radius error. The
Wronskian residual is primarily an implementation check and does not by itself
bound the asymptotic-tail error. Nevertheless, the positive gain below
$\omega=1/2$ and the negative gain above it are well separated from the
observed numerical variations.

\section{The Hasimoto transformation with all compatibility steps}
\label{sec:hasimoto}

\subsection{The geometric problem and its variables}

This section retains the geometric construction that motivated part of the
earlier argument, but specifies its domain. Let $\bm X(s,t)$ be a sufficiently
smooth space curve parametrized by arc length, and let
\begin{equation}
 \bm T=\partial_s\bm X,\quad|\bm T|=1,\quad
 \bm B=\bm T\times\bm N_{\!F}.
\end{equation}
Here $\bm N_{\!F}$ is the real Frenet normal. On a patch with curvature
$\kappa>0$, the Frenet--Serret equations are
\begin{equation}
 \bm T_s=\kappa\bm N_{\!F},\qquad
 (\bm N_{\!F})_s=-\kappa\bm T+\tau_g\bm B,\qquad
 \bm B_s=-\tau_g\bm N_{\!F}.
 \label{eq:frenet}
\end{equation}
The subscript $g$ distinguishes torsion $\tau_g$ from the acoustic time
coordinate $\tau$ used earlier. Assume binormal motion
\begin{equation}
 \bm X_t=\nu\,\bm X_s\times\bm X_{ss}
 =\nu\kappa\bm B,\qquad\nu>0.
\end{equation}
The coefficient $\nu$ has dimensions length squared per time. Introduce
$u=\nu t$, so that
\begin{equation}
 \bm X_u=\kappa\bm B.
 \label{eq:binormal}
\end{equation}
This is the specified curve evolution, not the Dirac equation.

\subsection{Complex frame and its spatial derivatives}

Define
\begin{equation}
 \phi(s,u)=\int_{s_0}^{s}\tau_g(s',u)\,\dd s',\qquad
 q_H=\kappa\ee^{\ii\phi},\qquad
 \bm N=(\bm N_{\!F}+\ii\bm B)\ee^{\ii\phi}.
 \label{eq:hasimoto-map}
\end{equation}
The dot product below is extended complex bilinearly; conjugation is written
explicitly. The orthogonality relations are
\begin{equation}
 \bm T\cdot\bm T=1,\quad\bm T\cdot\bm N=0,\quad
 \bm N\cdot\bm N=0,\quad\bm N\cdot\bm N^*=2.
 \label{eq:frame-products}
\end{equation}
For example,
$(\bm N_{\!F}+\ii\bm B)\cdot(\bm N_{\!F}+\ii\bm B)=1-1=0$,
whereas the conjugated product equals $1+1=2$.

Differentiating $\bm N$ and using $\phi_s=\tau_g$ gives
\begin{align}
 \bm N_s&=\left[-\kappa\bm T+\tau_g\bm B
 -\ii\tau_g\bm N_{\!F}
 +\ii\tau_g(\bm N_{\!F}+\ii\bm B)\right]\ee^{\ii\phi}
 \nonumber\\
 &=-\kappa\ee^{\ii\phi}\bm T=-q_H\bm T.
 \label{eq:complex-frame-space}
\end{align}
Conversely, the first Frenet equation becomes
\begin{equation}
 \bm T_s=\tfrac12(q_H^*\bm N+q_H\bm N^*),
 \label{eq:tangent-space}
\end{equation}
since the two terms in parentheses sum to $2\kappa\bm N_{\!F}$.

\subsection{Time derivatives from the binormal equation}

Because $\bm T=\bm X_s$, commuting the mixed derivatives in
Eq.~\eqref{eq:binormal} gives
\begin{equation}
 \bm T_u=(\kappa\bm B)_s
 =\kappa_s\bm B-\kappa\tau_g\bm N_{\!F}.
\end{equation}
Using
$(q_H)_s=(\kappa_s+\ii\kappa\tau_g)\ee^{\ii\phi}$, direct multiplication
shows
\begin{equation}
 (q_H)_s\bm N^*-(q_H)_s^*\bm N
 =-2\ii(\kappa_s\bm B-\kappa\tau_g\bm N_{\!F}).
\end{equation}
Consequently
\begin{equation}
 \bm T_u=\frac{\ii}{2}
 [(q_H)_s\bm N^*-(q_H)_s^*\bm N].
 \label{eq:tangent-time}
\end{equation}

Expand the remaining derivative in the complex frame:
\begin{equation}
 \bm N_u=\alpha\bm N+\beta\bm N^*+\gamma\bm T.
\end{equation}
Differentiating $\bm N\cdot\bm N=0$ gives
$0=2\bm N\cdot\bm N_u=4\beta$, hence $\beta=0$.
Differentiating $\bm N\cdot\bm N^*=2$ gives
$2(\alpha+\alpha^*)=0$, so $\alpha=\ii a$ with real $a(s,u)$.
Finally,
\begin{equation}
 0=(\bm T\cdot\bm N)_u
 =\bm T_u\cdot\bm N+\bm T\cdot\bm N_u
 =\ii(q_H)_s+\gamma.
\end{equation}
It follows that
\begin{equation}
 \boxed{\bm N_u=\ii a\bm N-\ii(q_H)_s\bm T.}
 \label{eq:complex-frame-time}
\end{equation}
The factor $\ii$ multiplying the real frame-rotation coefficient cannot be
dropped.

\subsection{Compatibility and the nonlinear Schr\"odinger equation}

Differentiate Eq.~\eqref{eq:complex-frame-space} with respect to $u$:
\begin{align}
 \bm N_{su}
 &=-(q_H)_u\bm T-q_H\bm T_u\nonumber\\
 &=-(q_H)_u\bm T
 +\tfrac{\ii}{2}q_H(q_H)_s^*\bm N
 -\tfrac{\ii}{2}q_H(q_H)_s\bm N^*.
 \label{eq:compatibility-one}
\end{align}
Differentiate Eq.~\eqref{eq:complex-frame-time} with respect to $s$:
\begin{align}
 \bm N_{us}
 &=\ii a_s\bm N+\ii a\bm N_s
 -\ii(q_H)_{ss}\bm T-\ii(q_H)_s\bm T_s\nonumber\\
 &=-\ii[a q_H+(q_H)_{ss}]\bm T
 +\left[\ii a_s-\tfrac{\ii}{2}(q_H)_sq_H^*\right]\bm N
 -\tfrac{\ii}{2}(q_H)_s q_H\bm N^*.
 \label{eq:compatibility-two}
\end{align}
Equate the coefficients in the independent frame
$(\bm T,\bm N,\bm N^*)$. The $\bm T$ coefficient gives
\begin{equation}
 \ii(q_H)_u+(q_H)_{ss}+a q_H=0.
\end{equation}
The $\bm N$ coefficient gives
\begin{equation}
 a_s=\tfrac12[q_H(q_H)_s^*+(q_H)_sq_H^*]
 =\tfrac12\partial_s|q_H|^2.
\end{equation}
The $\bm N^*$ coefficients already agree. Integrating in $s$ yields
\begin{equation}
 a(s,u)=\tfrac12|q_H|^2+A(u),
\end{equation}
where $A$ is real and independent of $s$. Thus
\begin{equation}
 \ii(q_H)_u+(q_H)_{ss}
 +\left[\tfrac12|q_H|^2+A(u)\right]q_H=0.
 \label{eq:hasimoto-with-phase}
\end{equation}
Set $q_H=Q_H\exp[\ii\int^u A(v)\dd v]$. Differentiating this phase
produces $-A Q_H$ in $\ii(q_H)_u$, canceling the $+A Q_H$ term.
The result is the focusing NLS equation
\begin{equation}
 \boxed{\ii(Q_H)_u+(Q_H)_{ss}+\tfrac12|Q_H|^2Q_H=0.}
 \label{eq:hasimoto-nls}
\end{equation}
This is the Hasimoto transformation \cite{Hasimoto1972}, now with explicit
complex conjugations, compatibility equations, and the removable time phase.

\subsection{What the geometric transformation does and does not prove}

Since $q_H$ is a curvature times a phase, $[q_H]=\mathrm{length}^{-1}$.
In contrast, the three-dimensional pair field is normalized as a number-density
amplitude and has $[\Psi_{\mathrm B}]=\mathrm{length}^{-3/2}$.
Their coordinates and time scalings are also different. Furthermore, the
repulsive pair equation with $g_{\mathrm B}>0$ is defocusing when written as
\begin{equation}
 \ii\hbar\partial_t\Psi_{\mathrm B}
 +\frac{\hbar^2}{2M_{\mathrm B}}\nabla^2\Psi_{\mathrm B}
 -g_{\mathrm B}|\Psi_{\mathrm B}|^2\Psi_{\mathrm B}
 -U_{\mathrm B}\Psi_{\mathrm B}=0,
\end{equation}
whereas Eq.~\eqref{eq:hasimoto-nls} has a focusing cubic sign.
Changing a positive normalization or rescaling real coordinates does not by
itself remove this sign distinction.

The function $A(u)$ is a frame-phase freedom; removing it cannot supply a
scattering energy reservoir. Thus the Hasimoto derivation establishes a
geometric correspondence for binormal curve evolution. To turn it into a
fermionic amplification model one would additionally need a microscopic
identification of the curve variables, a compatible effective action, a
background energy source, physical asymptotic channels, and their conserved
flux. Equations~\eqref{eq:gp}--\eqref{eq:collective-gain} provide a separate
conditional collective-wave route; they are not consequences of identifying
$Q_H$ with a Dirac spinor.

\section{Physical interpretation, prior results, and open problems}
\label{sec:discussion}

The three sectors can now be compared using their actual observables:
\begin{center}
\small
\begin{tabular}{>{\raggedright\arraybackslash}p{0.21\textwidth}
>{\raggedright\arraybackslash}p{0.28\textwidth}
>{\raggedright\arraybackslash}p{0.38\textwidth}}
\toprule
Sector & Conserved or constrained object & Supported statement\\
\midrule
Classical Dirac wave & Positive number density and conserved Dirac current &
One-sided passive scattering gives $\Refl_{\mathrm D}\leq1$.\\[3pt]
Quantized fermion field & Canonical anticommutators and state-dependent mode mixing &
$N_{\mathrm{out}}=|\beta|^2$ can be nonzero, with occupation limited by Pauli
blocking.\\[3pt]
Pair-condensate sound & Indefinite collective norm and acoustic energy flux &
The specified rotating absorbing background gives
$\Refl_{\mathrm{coll}}>1$ for $0<\omega<m\OH$ if transmission is nonzero.\\
\bottomrule
\end{tabular}
\end{center}

The collective mechanism is compatible with established condensate
superradiance calculations \cite{GiacomelliCarusotto2021a}, and with quantum
phonon emission in related Bose-condensate settings
\cite{GiacomelliCarusotto2021b}. Laboratory rotational amplification of water
waves supplies an experimental demonstration of the draining-vortex scattering
mechanism in a different medium \cite{Torres2017}; it is not evidence that
the present fermionic pair realization has already been achieved.

The boundary dependence is precise. Boundaries determine which channels are
occupied, whether negative collective norm is absorbed, whether a reservoir
feeds the system, and whether amplified waves return. They do not arbitrarily
change the bulk conservation law of a fixed Hermitian Dirac model. In a
single-mode feedback approximation, an outer amplitude reflection coefficient
$r_{\mathrm w}$ and an inner scattering amplitude $r_{\mathrm s}$ produce a
round-trip multiplier
\begin{equation}
 \mathcal M=r_{\mathrm w}r_{\mathrm s}\ee^{\ii\varphi_{\mathrm{rt}}}.
\end{equation}
Amplitude growth requires $|\mathcal M|>1$ together with the appropriate
round-trip phase condition. A single-pass gain does not automatically supply
either a reflecting wall or a resonance. This is why scattering
amplification and a superradiant instability must be analyzed separately.

The central unresolved task is microscopic realization: one should specify a
fermionic interaction and density, calculate pair parameters and pair-breaking
thresholds, construct a stationary sink-supported rotating solution, and solve
its full dispersive collective scattering problem. One should then compare
the reflected energy with the complete incident energy and the measured
background loss. A failure of pair stability, a missing negative-norm outgoing
channel, or a reflecting rather than absorbing core could invalidate the
collective amplification inferred from the ideal model. These are concrete
tests of the proposed interpretation.

The general principles used in this manuscript are not new discoveries.
The scope of the present revision is to make the earlier question
\cite{ChenZheng2023} mathematically well posed, correct the transfer of a
scalar identity to a Dirac system, and give an explicit conditional
pair-condensate scattering calculation. A claim of a new microscopic effect
or experimental proposal requires the further analysis just described.

\section{Conclusions}

For a minimally coupled classical Dirac field with real background potentials,
current conservation and positive number density constrain the reflected flux
under the stated one-sided passive boundary conditions. The massless
one-dimensional electrostatic system is exactly diagonal in its two
chiralities and has zero reflection. Massive step matching gives the same
non-amplification bound after the outgoing branch and velocity normalization
are chosen consistently. The scalar Wronskian identity differs because it
belongs to a different wave equation and a different norm.

Two broader statements remain possible and physically distinct. A quantized
fermion field can produce particles through nonzero Bogoliubov mixing, while
respecting the fermionic normalization and Pauli blocking. A system of paired
fermionic constituents can also support bosonic collective waves whose
effective norm is indefinite. Within a specified neutral dilute-pair theory
and nondispersive rotating absorbing-flow model, these waves obey the
superradiant condition $0<\omega<m\OH$. The numerical example reproduces this
conditional result without altering the classical Dirac flux law.

Finally, the Hasimoto transformation maps binormal curve dynamics to a
focusing NLS equation. Its intermediate steps do not establish an identity
between curvature, a Dirac spinor, and a repulsive pair-condensate order
parameter. Maintaining these distinctions gives a consistent basis for
investigating superradiant phenomena in systems that contain fermions.

\appendix
\section{Uncertainty relations and why boundary selection does not violate them}
\label{app:uncertainty}

The distinction between preparation uncertainty and measurement disturbance
is relevant because the earlier discussion invoked a neutron measurement
experiment. Let $A'=A-\langle A\rangle$ and
$B'=B-\langle B\rangle$ for a normalized state in the common domains required
for the products below. Cauchy--Schwarz gives
\begin{equation}
 (\Delta A)^2(\Delta B)^2
 =\|A'\psi\|^2\|B'\psi\|^2
 \geq|\langle A'\psi,B'\psi\rangle|^2.
\end{equation}
The magnitude squared of a complex number is at least the square of its
imaginary part. Since
\begin{equation}
 \operatorname{Im}\langle A'\psi,B'\psi\rangle
 =\frac1{2\ii}\langle[A,B]\rangle,
\end{equation}
we obtain the Robertson relation \cite{Robertson1929}:
\begin{equation}
 \boxed{\Delta A\,\Delta B\geq\tfrac12
 |\langle[A,B]\rangle|.}
\end{equation}
For a canonical position-momentum pair on the appropriate common domain,
$[x,p]=\ii\hbar$ and $\Delta x\,\Delta p\geq\hbar/2$.
Projecting or conditioning a state changes the state, but the same derivation
applies to its normalized conditional state if the required domains and
finite variances exist. One cannot retain an unrestricted momentum
commutator while ignoring domain changes created by a boundary.

Measurement noise $\epsilon(A)$ and disturbance $\eta(B)$ are different
quantities from the initial-state standard deviations. In the measurement
framework considered by Ozawa, the corresponding relation is
\begin{equation}
 \epsilon(A)\eta(B)+\epsilon(A)\Delta B+\Delta A\eta(B)
 \geq\tfrac12|\langle[A,B]\rangle|
 \label{eq:ozawa}
\end{equation}
\cite{Ozawa2003}. The neutron spin experiment in Ref.~\cite{Erhart2012}
concerns an error--disturbance relation. It is not a demonstration that
the canonical preparation inequality fails. Neither relation supplies an
extra term in Eq.~\eqref{eq:continuity}; such a term must follow from the
actual equation of motion or an explicitly coupled environment.

\section{A direct coefficient check for the focusing NLS}
\label{app:soliton}

An independent check of the cubic coefficient in
Eq.~\eqref{eq:hasimoto-nls} is furnished by
\begin{equation}
 Q_H=2\eta\,\operatorname{sech}[\eta(s-2bu)]
 \exp\{\ii[bs+(\eta^2-b^2)u]\},\qquad\eta>0,
 \label{eq:nls-soliton}
\end{equation}
where $b$ is a real wave-number parameter. Let
$\zeta=s-2bu$, $f(\zeta)=2\eta\operatorname{sech}(\eta\zeta)$, and
$\Theta=bs+(\eta^2-b^2)u$. Then
\begin{align}
 (Q_H)_u&=\ee^{\ii\Theta}
 [-2bf'+\ii(\eta^2-b^2)f],\\
 (Q_H)_{ss}&=\ee^{\ii\Theta}[f''+2\ii bf'-b^2f].
\end{align}
Thus
\begin{equation}
 \ii(Q_H)_u+(Q_H)_{ss}+\tfrac12|Q_H|^2Q_H
 =\ee^{\ii\Theta}[f''-\eta^2f+\tfrac12f^3].
\end{equation}
Differentiating $\operatorname{sech}$ twice gives
$f''=\eta^2f-\tfrac12f^3$, so the residual vanishes.
Here $[\eta]=[b]=\mathrm{length}^{-1}$ and $[u]=\mathrm{length}^2$,
consistent with the geometric time scaling. This exact solution checks the
NLS coefficient; it does not identify its field with the pair density.

\end{document}